\documentstyle[prl,aps]{revtex}
 \twocolumn
 
\begin{document}
 \draft
\title{Complexity as the driving force for dynamical glassy transitions}

 \author{Th.~M.~Nieuwenhuizen}
 \address{
 Van der Waals-Zeeman Laboratorium, Universiteit van Amsterdam\\
	 Valckenierstraat 65, 1018 XE Amsterdam, The Netherlands\\}
 \date{March 20, 1995; email: nieuwenh@phys.uva.nl}
\maketitle
\begin{abstract}
Aspects of the dynamical glass transition are considered within a
mean field spin glass model. At the dynamical transition the
the system condenses in a state of lower entropy.
The difference, the information entropy or complexity,
is calculated by analysis of the
metastable (TAP) states. It increases for lower temperatures, showing that
more and more free energy barriers cannot be overcome on
accessible times scales. Near $T_c$ the total glassy free
energy lies below the paramagnetic one, which makes the
transition unavoidable from a thermodynamic point of view.
The multitude of glassy states implies an extensive
 difference between the average specific heat
and the derivative of the average internal energy.
\end{abstract}
\pacs{64.70.Pf, 75.10Nr,75.40Cx,75.50Lk}
\narrowtext

 The structural glass transition is said to occur at
 the temperature $T_g$ where the viscosity equals $10^{14}$ Poise.
 The precise nature of this transition is not fully understood.
 Experimentally one often determines the entropy
 from the specific heat data by integrating
 $C/T$ from a reference temperature in the liquid phase down to $T$.
 At zero temperature a residual entropy is found,
 in contradiction with
 the third law of thermodynamics.

 Alternatively, a glass can be seen as a disordered solid.
 In this description the glass transition is a transition to a
 state with much smaller entropy than the liquid.
 As the  free energy then becomes larger, it is not immediately clear from
 thermodynamic considerations why the system gets captured in such a state.
 The answer is that the condensed system then has lost  the
 entropy of selecting one out of the many
 equivalent states; this part of the entropy is called the
 {\it information entropy} or
 {\it complexity} ${\cal I}$.~\cite{Jackle}~\cite{Palmer}
 Its origin can be understood as follows.
 When the relevant state $a$ has a large degeneracy ${\cal N}_a\equiv
 \exp({\cal I}_a)$, the
 partition sum yields $Z=\sum_b\exp(-\beta F_b)\approx
{\cal N}_a\exp(-\beta F_a)$, so
 $F=F_a-T{\cal I}_a$ is the full free energy of the system.
 The entropy loss arises when the system chooses the state to condense into,
 since from then on only that single state is observed.~\cite{quantummeas}
 As the total entropy $S=S_a+{\cal I}_a$ is continuous,
 so is the total free energy. If the weights $p_a$ of the states $a$ are
 fixed at the transition,\cite{Jackle} one can assume that the free energy
 difference between the condensed phase and the liquid is positive and
 grows quadratically below $T_c$.
 This explains the well known discontinuity in quantities such as the
 specific heat. However, it is a bit unsatisfactory that
 this higher free energy branch describes the physical state.

  Recent work on glasses mainly has been performed from a dynamic approach.
 Mode coupling equations starting from the liquid phase have been studied
 for some while.\cite{Gotze} At a critical
 temperature $T_c>T_g$ a dynamic phase transition has been reported.
 The presence of this transition has been questioned, however.\cite{Rudi}

 We wish to analyze these and related questions within a
relatively well understood spin glass model, the mean
 field $p$-spin interaction spin glass, defined by the Hamiltonian
 (\ref{Ham=}).
 Kirkpatrick and Thirumalai\cite{KirkpT} pointed out that for
 the case $p=3$ there is a close analogy with
 models for the structural glass transition. Moreover, its properties are
 quite insensitive of the value of $p$ as long as $p>2$.
 The spherical limit of this model has received some attention recently.
 Its static version was solved by Crisanti and Sommers.~\cite{CS}
 They find a static first order transition to a state with
 one step replica symmetry breaking (1RSB) at a temperature $T_g$.
 The dynamics of this system was studied by Crisanti, Horner and Sommers
 (CHS)~ \cite{CHS} and Cugliandolo and Kurchan~\cite{CK}.
 Both groups find a dynamical transition at a temperature $T_c>T_g$,
 which can be interpreted on a quasi-static level as a 1RSB transition.
 For $T<T_c$ one of the fluctuation modes is massless (``marginal''),
 not unexpected for a glassy state.
At $T_c^-$ there is a drop in the specific heat.
 Integrating  $C/T$ to define the ``dynamical''  entropy
 yields a free energy that exceeds  the paramagnetic one  quadratically.
 Ref. \cite{KPV} discusses the interpretation of metastable
 states (``TAP-states'') in this system. The statistics of those states
 was  considered by Crisanti and Sommers (CS).~\cite{CSTAP}
 They reproduced the dynamical free energy obtained  in \cite{CHS}.
 As a side remark, let us draw attention to the model
 with  both random pair couplings and random  quartet couplings.
 It is an exactly solvable spin glass model with infinite order replica
 symmetry breaking,~\cite{Nqsg} which  remains exactly solvable after
  quantization of the spherical spins.~\cite{Nqsm}

The present author recently proposed a replica calculation as a
short cut for the long time result of the dynamical analysis.~\cite{maxmin}
He took the principle point of view that a dynamical transition
will automatically  get trapped in a state with diverging time scale, if
present. In a replica analysis this led him to replace Parisi's stationarity
criterion for the location of the breakpoint of the 1RSB solution by a
marginality criterion. In this way the correct dynamical
transition temperature and shape of the order parameter function were
reproduced from replica's for a number of model systems.
 However, as the free energy is minimized in this procedure, it was found
to lie  near $T_c$ {\it below} the paramagnetic value and to have
a larger slope.
Though this is exactly what one expects at a first order phase transition,
such a replica result cannot be trusted without independent verification
and interpretation.

 For a system with $N$ spins
we consider the Hamiltonian
\begin{equation}\label{Ham=}
{\cal H}=-\sum_{i_1<i_2<\cdots<i_p} J_{i_1 i_2\cdots i_p}
S_{i_1}S_{i_2}\cdots S_{i_p}-H\sum_iS_i
\end{equation}
with independent Gaussian random couplings, that have average zero
and variance $J^2p!/2N^{p-1}$. The spins are subject to the spherical
condition $\sum_i S_i^2=N$.
At zero field the replica calculation gives the free energy~\cite{CS}
\begin{eqnarray}\label{bFCS}
\frac{ F}{N}&=&-\frac{\beta J^2}{4}+\frac{\beta J^2}{4}
\xi q^p\\
&-&\frac{T}{2x}\log(1-\xi q)+\frac{T\xi}{2x}\log(1-q)
\nonumber\end{eqnarray}
where $\xi=1-x$. The first term describes the paramagnetic free energy.
Here and in the sequel, we omit the $T=\infty$ entropy. It is a constant,
only fixed after quantizing the spherical
model,~\cite{Nqsm} that plays no role
in the present discussion.
For the marginal solution $q$ is fixed by equating the lowest
fluctuation eigenvalue to zero, which gives
$p(p-1)\beta^2 J^2q^{p-2}(1-q)^2/2=1$. The condition
$\partial F/\partial q=0$ then yields $x=x(q)\equiv(p-2)(1-q)/q$.
This dynamical transition sets in at  temperature
$T_{c}=J \{p(p-2)^{p-2}/2(p-1)^{p-1}\}^{1/2}$ where $x$ comes below
unity.

The metastable states of this system, labeled by
$a=1,2,\cdots,{\cal N}$, have local magnetizations
$m_i^a=\langle S_i\rangle^a$. At $H=0$ their free energy
$F_a=F_{TAP}(m_i^a)$ is a minimum of the
``TAP'' free energy ~\cite{KPV}
\begin{eqnarray} \label{FTAP=}
&F&_{TAP}(m_i)=
-\sum_{i_1<i_2<\ldots<i_p}J_{i_1 i_2\cdots <i_p}
m_{i_1}m_{i_2}\cdots m_{i_p} \nonumber\\
&-&\frac{NT}{2}\log(1-q)
-\frac{N\beta J^2}{4}(1+(p-1)q^p-pq^{p-1})
\end{eqnarray}
where now $q=(1/N)\sum_i m_i^2$ is the self-overlap.
Palmer ~\cite{Palmer} calls the metastable states ``components''.
According to the Gibbs weight their probability of occurrence is
$p_a=\exp(-\beta F_a(T))/Z$, with $Z=\sum_a\exp(-\beta F_a)$.
Given the $p_a$'s one can calculate the
``component averages'' such as $\overline F=\sum_a p_aF_a$ and
 $\overline S=\sum_a p_aS_a$. For a given sample they are the objects one
finds when repeating the measurement very often,
and averaging over the outcomes.

The nice thing of the present model is that many questions can be answered
directly. After setting $\partial F_{TAP}/\partial m_i=0$, we can use
this equation to express $F_a$ in terms of $q_a$
alone. This gives $F_a=Nf(q_a)$ where
\begin{eqnarray}\label{fiTAP=}
f(q)&=&\frac{\beta J^2}{4}[-1+(p-1)q^p-(p-2)q^{p-1}]\nonumber\\
&-&\frac{Tq}{p(1-q)}
-\frac{T}{2}\log(1-q)
\end{eqnarray}
Since $F_a$ only depends on the selfoverlap $q_a$, it
is self-averaging. In the paramagnet one has $m_i=q=0$, so both eqs.
(\ref{FTAP=}) and (\ref{fiTAP=}) reproduce the replica free energy
$F/N=-\beta J^2/4$.  From the replica analysis we know that at
 $T_c^-$ the value of $q$ is $q_c=(p-2)/(p-1)$.
The component free energy $F_a=Nf(q_c)$
exceeds the free energy $-\beta J^2N/4$ of the
paramagnet. As expected from experimental knowledge on glasses,
the internal energy is continuous.
The difference is solely due to the lower entropy,
$S_a=-N\beta^2 J^2/4-{\cal I}_c$, where
\begin{equation} \label{I=}
{\cal I}_c=N\left(\frac{1}{2}\log(p-1)+\frac{2}{p}-1\right)
\end{equation}
is the value of complexity at the transition point.

This discussion shows explicitly that the entropy of the component
the system condenses in,
as well as its average $\overline S$,
is much smaller that the entropy of the paramagnet.
In the quantized system ${\overline S}$ will vanish at $T=0$.~\cite{Nqsg}

A simple calculation shows that
the dynamical free energy of CHS and CS, and the marginal replica free energy
obtained from eq. (\ref{bFCS})~\cite{maxmin} have the following
connection with the component average free energy ${\overline F}=Nf(q)$:
\begin{eqnarray}\label{fdynfeff}
F_{dyn}&=&{\overline F}-T{\cal I}_c\\
F&=&{\overline F}-\frac{T{\cal I}_c }{x(q)}
\label{Frepl}
\end{eqnarray}
Since $x=1$ at the transition, both are continuous there.

Due to the difference between
(\ref{Frepl}) and
(\ref{fdynfeff}) we have decided to redo the analysis
of the TAP equations. Hereto we consider the generalized partition sum
\begin{equation}\label{Zu=}
Z_u=\sum_a e^{-u\beta F_a(T)}
\end{equation}
 For $u=0$ we thus calculate the
total number ${\cal N}$ of TAP-states, while for $u=1$ we consider their
partition sum.
The sum over the TAP states can be calculated using standard approaches.
\cite{BM}
One goes to integrals over all $m_i$-values of
$\exp(-u\beta F_{TAP}(m_i))$
$\prod_i\delta\left(\partial \beta F_{TAP}/\partial m_i\right)$det$
\left(\partial^2 \beta F_{TAP}/\partial m_i\partial m_j\right)$~\cite{absval}
and introduces conjugate fields for writing the delta functions and the
determinant as exponential integrals.
After averaging over disorder, bilinear overlaps of these field
act as order parameters at a saddle point.
They are introduced at the cost of conjugated order parameters, that we
indicate by a hat. The replicated free energy reads
\begin{equation}\label{Fun}
\frac{ F_u^{(n)}}{N}=
u\sum_{\alpha=1}^n f(q_{\alpha\alpha})-T\Phi_n
\end{equation}
\begin{eqnarray}
\Phi_n&=&\sum_{\alpha\beta}\left\{
q_{\alpha\beta}\hat q_{\alpha\beta}
+y_{\alpha\beta}\hat y_{\alpha\beta}
+z_{\alpha\beta}\hat z_{\alpha\beta}
+\frac{\lambda_{\alpha\beta}}{4}
(y_{\alpha\beta}^2-z_{\alpha\beta}^2)\right\}\nonumber\\
&+&{\rm tr}\ln (a-\hat z)
-\frac{1}{2}{\rm tr} \ln\left[(a-\hat y)^2+\sigma \hat q \right]\end{eqnarray}
Here $\lambda_{\alpha\beta}=\beta^2 J^2p(p-1)q_{\alpha\beta}^{p-2}$,
$a_{\alpha\beta}=a_\alpha\delta_{\alpha\beta}$ with
$a_\alpha\equiv\lambda_{\alpha\alpha}
(1-q_{\alpha\alpha})/2+1/(1-q_{\alpha\alpha})$ and
$\sigma_{\alpha\beta}=p\beta^2 J^2 (q_{\alpha\beta})^{p-1}$.
Usually one assumes that no replica symmetry breaking occurs when one
calculates the number of TAP states. However, since in eq. (\ref{Fun})
it does occur at $u=1$, it is not a well motivated assumption that it
would not occur for $u=0$.
Having in mind  the 1RSB solution of the ordinary partition sum,
we are therefore led to investigate a similar breaking
scheme in the replicated TAP approach. We thus set $q_{\alpha\beta}\equiv
\langle m_\alpha m_\beta \rangle=(q_0-q_1)\delta_{\alpha\beta}+
q_1{\cal E}_{\alpha\beta}$, where ${\cal E}$ is Parisi's
1RSB matrix, having ${\cal E}=1$ on all $n/x$ diagonal blocks
of size $x\ast x$, while zero elsewhere.
A similar 1RSB pattern is assumed for
the 5 other order parameters. At fixed $x$ the 12-dimensional saddle point
can be found explicitly; details will be given elsewhere.
As expected, the above replica expression for $q$ is found back as
 solution of $\partial f(q_0)/\partial q_0=0$ at $q_0=q$.
The result $q_1=q_0$ asserts that the mutual overlap
between different states in the same cluster is equal to the selfoverlap.
As in the replica calculation of the ordinary partition sum,
$x$ can still take any value.
For the long time limit of the
dynamical approach the marginality condition
should be taken,~\cite{maxmin} in the form given below eq. (\ref{bFCS}).
It turns out that the fluctuation matrix then automatically has 3
zero eigenvalues, proving the marginality.
The lowest acceptable value of $F_u(T;x)$ is obtained when $x$
takes the boundary value
\begin{equation}\label{x=u}
x(q,u)=\frac{x(q)}{u}=(p-2)\frac{1-q}{uq}
\end{equation}
where one more eigenvalue vanishes.
At the marginal point one then finds
 $\Phi_n=n{\cal I}_c/Nx(q,u)$, which implies
\begin{equation}\label{fuans}
F_u(T)=u\left[{\overline F}(T)-\frac{T{\cal I}_c }{ux(q,u)}
\right]=uF(T)
\end{equation}
For $u=1$ we thus recover the replica result (\ref{Frepl}) rather than
the dynamical expression (\ref{fdynfeff}). The latter expression is
still correct in the sense of CHS, namely as the component average
free energy with entropy shifted over the constant ${\cal I}_c$,
so that it is continous at $T_c$. However, after quantization
 ${\cal I}_c$ then remains as residual entropy at $T=0$.
The derivation of eq. (\ref{fdynfeff}) by
CS is based on two assumptions that are both subject to criticism.

 From eq. (\ref{Frepl}) we observe that the complexity is temperature
dependent: ${\cal I}(T)={\cal I}_c/x(q(T))$. Its increase towards
smaller $T$
shows that more and more components become available for the system.
In other words, when decreasing the temperature phase space splits
up more and more. At each temperature there appear new
free energy barriers which can no longer be overcome.
This well known notion of mean field spin glass theory, which nicely
explains temperature cycling aging experiments,~\cite{Vincent}
is unknown in glass theory.
Note that ${\cal I}(T)$ diverges as $1/T$ at low $T$.
This leads to an infinite complexity at $T=0$
for time scales where the marginal state describes the physics.
These will probably include the experimentally accessible time scales.

It would be nice to have a measurable quantity that probes
the multitude of solutions. One object that should be accessible, at least
numerically, is the specific heat. The standard expression
$C=dU/dT=\sum_a d(p_aU_a)/dT$ is likely to differ from the component
average ${\overline C}=\sum_a p_a dU_a/dT$.~\cite{Palmer} The interesting
question is whether their difference is extensive.
Based on experience in a toy model,~\cite{NvR} we think it
generally is. Since in the present model
the energy fluctuations are too small at $H=0$,~\cite{fef}
 it can only occur in a field.
{}From the CHS or replica internal energy $U$ in a small field
we obtain
\begin{eqnarray}
\frac{1}{N}C(T,H)&=&\frac{1}{2}\beta^2J^2(1+(p-1)q^p-pq^{p-1})\nonumber\\
&-&\beta^2H^2\frac{(p-1)^2(p-2)(1-q)^2}{p(pq+2-p)}\end{eqnarray}
On the other hand, a short calculation shows that $C_a=TdS_a/dT$
remains only a function of $q_a$ at the marginal point, which
in the present model is field-independent. This
implies that ${\overline C}$ is field-independent as well,
thus satisfying the Parisi-Toulouse hypothesis~\cite{PaT}. Interestingly
enough, we find $C<{\overline C}$, whereas Palmer derives the
opposite. Our reversed ``dynamical'' inequality is a new
result that is due to the marginality.

Some more aspects are of interest. Coming from the low temperature side,
the transition has the nature of a continuous glassy transition.
In a replica formulation states with breakpoint $x=x(q)$
have massive longitudinal fluctuations ($\Lambda'\sim T_c-T$)
and massless transversal fluctuations. (The latter modes
are commonly called the ``replicon'';
we have proposed the more physical name ``ergodon''~\cite{Nqsg}).
This structure is very reminiscent of
the behavior of fluctuations in the SK model, be it much simpler.
In an external field the degeneracy of the longitudinal eigenmode is
lifted. It remains as a non-degenerate small eigenvalue which
vanishes at the same point where the transversal
``ergodon'' becomes marginal.
Therefore the same picture remains valid in an external field.
 In particular,
there is a de Almeida-Thouless line where the paramagnet becomes
unstable with respect to non-linear perturbations related to the
1RSB solution discussed.

It would be useful to study the loop expansion.
This may shed light on the lower critical dimension
of the problem and on the relation with structural glasses.

In finite (but large?) dimensions
our explanation is incompatible with a droplet picture.~\cite{droplet}
The assumption that the condensed phase is non-degenerate
implies ${\cal I}_c/N=0$. Then it remains unclear from
a thermodynamic point of view why a transition to a state with
higher free energy can occur.
It is also not obvious whether a droplet approach can explain
 our reversed inequality $C<{\overline C}$.

In conclusion, we have considered a mean field model for a spin glass.
As it undergoes a first order phase transition to a state with
one step of replica symmetry breaking, it
has many features of models for glassy transitions, possibly in the
absence of disorder.  Within this model we have found that the
following scenario takes place:

\noindent
- At $T_c$ the system freezes in a metastable
state with free energy much higher than that of the paramagnet.

\noindent
- The component average entropy ${\overline S}$ drops at $T_c$
by an extensive amount ${\cal I}_c$, related to the degeneracy of the
glassy states at $T_c$.

\noindent
- The entropy $S={\overline S}+{\cal I}$, the internal energy,
and the free energy are continuous.

\noindent
- ${\overline S}$ vanishes at $T=0$, so there is no residual
entropy.

\noindent
- Below $T_c$ the glassy free energy lies below the paramagnetic one,
and has a larger slope. Therefore the transition is first order
and unavoidable in the thermodynamic sense.
The driving force is the complexity.

\noindent
- There is no latent heat, but a there is a drop in the heat capacity.

\noindent
- In an adiabatic cooling experiment the weights of the states are
initially set at $T_c$. However, there will be a further bifurcation
of phase space at any lower $T$.

\noindent
- The inequality $C/N<{\overline C}/N$ is a direct result of the multitude
of states and their marginality.

Though having considered only one specific model,
we are tempted to believe that the developed picture is very general.
The role of disorder is probably not essential.
What is needed is an extensive complexity. It seems to occur at many glassy
and glass transitions. However,
continuous transitions with infinite
order replica symmetry breaking are known to behave differently.

\acknowledgments
The author thanks
J. Kurchan, H. Rieger, R. Schmitz and P. de Ch\^atel for discussion,
J.M. Luck for performing the stability analysis with Macsyma, and
H.J. Sommers for communicating ref. ~\cite{CSTAP} prior to publication.
He is grateful for hospitality at the S.Ph.T. of the CEA Saclay,
where part of this work was performed.
This work was made possible by the Royal Dutch Academy of Arts
and Sciences (KNAW).

2) As CS approximated $[\ln {\cal N}_a]_{av}$ by $\ln[{\cal N}_a]_{av}$,
we can follow their steps by neglecting replica symmetry breaking
through putting $x\equiv 1$.
We then find the same saddle point as function of $q$,
and $\Phi_n=n{\cal I}_c/N$ reproduces eq. (\ref{fdynfeff}).
However, in this calculation one of
the fluctuation modes becomes unstable when $u$
becomes less than $(p-2)(1-q)/q$; this is the remnant of eq. (\ref{x=u})
for $x(q,u)=1$. At $u=0$ it thus proves that the CS saddle point
is unstable w.r.t. replica symmetry breaking.
\bibitem{Vincent} F. Lefloch, J. Hammann, M. Ocio, and E. Vincent,
Europhys. Lett. {\bf 18} (1992) 647
\bibitem{PaT} G. Parisi and G. Toulouse, J. Phys. (Paris) (1980)
\bibitem{fef} This happens since at $H=0$ its (free) energy fluctuations are
not ${\cal O}(\sqrt{N})$ but ${\cal O}(1)$, as can be seen
by expanding the result for $\ln [Z^n]_{av}$ to order $n^2$.
For $H=0$ there appear no terms of order $n^2N$; for $H\neq 0$ they
do appear.

\end{document}